\documentclass{article}

\usepackage{graphicx}
\usepackage[round]{natbib}
\usepackage{epsfig}

\title
{
Uncertain Climate Forecasts From Multimodel Ensembles: When to Use Them and When to Ignore Them
}

\author{Stephen Jewson\footnote{\emph{Correspondence email}: \texttt{stephen.jewson@rms.com}} and Ed Hawkins\\}

\begin{document}

\maketitle

\begin{abstract}
Uncertainty around multimodel ensemble forecasts of changes in future climate reduces the accuracy of those forecasts.
For very uncertain forecasts this effect may mean that the forecasts should not be used.
We investigate the use of the well-known Bayesian Information Criterion (BIC) to make the decision as to whether a forecast should be used or ignored.
\end{abstract}

\section{Introduction}

The climate predictions produced by numerical models predict \emph{changes} in climate relative to the present day,
rather than predicting absolute levels of future climate.
The ratio $r$ of the size of the change to the uncertainty
around that change is then a measure of the confidence one can have in a forecast.
It also determines whether or not it would be better to ignore a forecast, use it, or adjust it in some way.
We have studied this question in~\citet{jewsonh09c},
in which we compared the relative effectiveness of five different ways of using a multimodel numerical model climate forecast,
as a function of this ratio $r$.
We tested the different methods in terms of the mean squared error of predictions they would produce, using monte carlo simulations.
We found that for an ensemble based on 10 truly independent climate models,
for small values of this ratio ($r<0.74$) it would be better to ignore the forecast completely.
For larger values of this ratio ($0.73<r<2.02$) it is best to adjust the forecast towards zero,
and for the largest values of this ratio ($r>2.02$) it is best to use the forecast unadjusted.

These results cannot, however, be used literally as a decision rule.
This is because, in practice, we never know the exact value of the ratio $r$
since it is defined in terms of the \emph{actual} mean change in climate and the \emph{actual} uncertainty around that change
(which are unknown) rather than the \emph{estimated} change and the \emph{estimated} uncertainty (which are known).

It is tempting to replace the real values of the mean and uncertainty with their estimates,
calculate an estimate of $r$, and then use the ranges given above.
However, rather than do that it may be better to treat this as a statistical model selection problem,
and use standard model selection methods.
Perhaps the most standard model selection rule is the Bayesian Information Criterion (BIC)
(see many statistics textbooks, such as~\citet{wasserman}),
which gives a way to select between models which have been fitted by maximum likelihood.
Two of the five methods used in~\citet{jewsonh09c}
were fitted using maximum likelihood (those that involve either ignoring or using the ensemble mean)
with the others using a second stage of fitting for the `damping' parameters.
As a first step, therefore, we apply BIC to the two methods from~\citet{jewsonh09c} that were fitted using maximum likelihood, and postpone
to a future study the question of how to apply model selection to the remaining three methods.

In section~\ref{s2} we present the mathematical setup and the two different methods we will consider for turning
multimodel ensemble output into a probabilistic prediction.
In section~\ref{s3} we derive the BIC for each of the methods.
In section~\ref{s4} we compare the derived BIC values, and derive an expression for the critical value of $r$ that determines
the threshold between the two methods.
In section~\ref{s5} we apply BIC-based model selection to CMIP derived predictions of UK temperatures, and
finally in section~\ref{s6} we discuss the results.

\section{Two Methods for Interpreting Multimodel Numerical Model Climate Forecasts}
\label{s2}

Our mathematical setup follows that of~\citet{jewsonh09b} and~\citet{jewsonh09c}.
We imagine that we have an ensemble of $n$ climate models.
Ideally each climate model has been run in a large initial condition ensemble, although
in practice that may not matter if the impact of initial condition uncertainty is small
relative to model uncertainty.

We use a classical statistical framework in which we consider that the models
have the same error statistics and are samples from
an infinite population of all possible models with those error statistics.
We also assume, perhaps somewhat optimistically, that the mean of this population is equivalent to the true future climate
(we call this the `perfect ensemble assumption').
We suppose that the variance of this ensemble has been adjusted to compensate for possible dependencies between the models
(perhaps using the method of~\citet{jewsonh09a}) and hence that, because of this adjustment,
we can consider the adjusted ensemble members as being independent.

We write the mean of this ensemble as $m$ and the sample variance as $s^2$, where:
\begin{eqnarray}
m  &=&\frac{1}{n} \sum_{i=1}^{n} x_i\\
s^2&=&\frac{1}{n} \sum_{i=1}^{n} (x_i-m)^2
\end{eqnarray}

The uncertainty around the estimate of the ensemble mean can be estimated by $V$, where:
\begin{equation}
V=\frac{s^2}{n}
\end{equation}

We then consider the following two methods for turning this ensemble into a forecast:

\subsection{Use Ensemble Mean, Use Ensemble Variance}\label{m1}

In this method we make a prediction by fitting a normal distribution to the ensemble using maximum likelihood.
The mean and variance ($\mu_1,\sigma_1^2$) of the prediction are then given by:

\begin{eqnarray}
\hat{\mu}_1     &=&m\\
\hat{\sigma}_1^2&=&\frac{1}{n}\sum_{i=1}^{n} (x_i-\mu_1)^2\\
                &=&s^2
\end{eqnarray}

Note that we deliberately do not use the well known `$n-1$' expression for the variance,
since we will need our estimates of the parameters to be maximum likelihood estimates for BIC to apply.

\subsection{Ignore Ensemble Mean, Use Ensemble Variance}\label{m2}

In this method we fit a normal distribution to the ensemble using maximum likelihood, but with the assumption
that the mean of the distribution is zero.
In other words, we ignore the mean of the ensemble, but use the variance.

The mean and variance ($\mu_2,\sigma_2^2$) of the prediction are then given by:

\begin{eqnarray}
\hat{\mu}_2&=&0\\
\hat{\sigma}_1^2
&=&\frac{1}{n} \sum_{i=1}^{n} (x_i-0)^2\\
&=&\frac{1}{n} \sum_{i=1}^{n} (x_i-m+m)^2\\
&=&\frac{1}{n} \left( \sum_{i=1}^{n} (x_i-m)^2 + \sum_{i=1}^{n} m^2 \right)\\
&=&\frac{1}{n} \left( n s^2 + n m^2 \right)\\
&=&s^2+m^2
\end{eqnarray}

We do not consider forecasts in which we also ignore the variance of the ensemble,
or in which we use the mean but ignore the variance,
since doing so would be to assume that the
predicted mean change had no uncertainty at all.
That would not correspond to our beliefs,
and would also give a BIC score of infinity which would be beaten by the other two methods by default.

\section{BIC for each of the models}
\label{s3}

In this section we now derive BIC values for the two methods defined in section~\ref{s2}.

The definition of BIC is:
\begin{equation}\label{eq1}
B=-2l+k \ln n
\end{equation}

where $l$ is the likelihood attained at the maximum,
$k$ is the number of parameters in the model
and $n$ is the number of independent data points used to fit the model (see, for example, ~\citet{wasserman}).
When comparing a number of models, the model with the lowest value for the BIC is considered the best model.

As the number of parameters increases models will generally fit the data better and $l$ will increase.
Hence the $-2l$ term will decrease while the $k \ln n$ term will increase.
Whether the BIC itself increases or decreases depends on whether the extra parameters offer sufficient
additional explanatory power to improve predictions, or whether the model becomes overfitted.

Since we have forced independence between the different climate models by inflating the variance
the likelihood is given by the density of the multivariate normal distribution for independent
samples, which is:
\begin{equation}
p(x)=\prod_{i=1}^{n} \frac{1}{\sqrt{2\pi}} \frac{1}{\sigma} \mbox{exp} \left( -\frac{1}{2\sigma^2}(x_i-\mu)^2 \right)
\end{equation}

The log-likelihood is then:
\begin{eqnarray}
l(x)
&=&\ln p(x)\\
&=&\sum_{i=1}^{n} \ln (\frac{1}{\sqrt{2\pi}})+\sum_{i=1}^{n} \ln \frac{1}{\sigma} -\sum_{i=1}^{n} \frac{1}{2\sigma^2}(x_i-\mu)^2\\
&=&-\frac{n}{2}\ln 2\pi-\frac{n}{2} \ln \sigma^2 -\sum_{i=1}^{n} \frac{1}{2\sigma^2}(x_i-\mu)^2
\end{eqnarray}

Using this, we now derive expressions for the BIC for each model.
The differences in BIC between our two methods arise simply because of the different choices of $\mu$ and $\sigma$
given in sections~\ref{m1} and~\ref{m2} above.

\subsection{BIC for Use Ensemble Mean, Use Ensemble Variance}

For this method the log-likelihood is:
\begin{eqnarray}
l(x)
&=&-\frac{n}{2}\ln 2\pi-\frac{n}{2} \ln s^2 -\frac{n}{2}\\
&=&-\frac{n}{2} \left [\ln 2\pi+ \ln s^2+n\right]
\end{eqnarray}

and there are two estimated parameters $(k=2)$, and so, from equation~\ref{eq1}, the BIC is:

\begin{eqnarray}
B_1
&=& n \left [\ln 2\pi+ \ln s^2+n\right]+2 \ln n
\end{eqnarray}

\subsection{BIC for Ignore Ensemble Mean, Use Ensemble Variance}

For this method the log-likelihood is:
\begin{eqnarray}
l(x)
&=&-\frac{n}{2}\ln 2\pi-\frac{n}{2} \ln (s^2+m^2) -\frac{n}{2}\\
&=&-\frac{n}{2} \left [\ln 2\pi+ \ln (s^2+m^2)+n\right]
\end{eqnarray}

Note that the only difference from the analogous expression for the previous model is the $m^2$ term,
which is due to the bias in the forecast introduced by ignoring the ensemble mean,
and which drives the BIC upwards, and tends to make the forecast less favoured.

This time there is only one estimated parameter $(k=1)$, which, on the other hand,
tends to makes the forecast \emph{more} favoured, and so the BIC is:

\begin{eqnarray}
B_2
&=& n \left [\ln 2\pi+ \ln (s^2+m^2)+n\right] +\ln n
\end{eqnarray}

\section{Comparing BIC values between the two models}
\label{s4}

The difference between the BIC scores for our two methods is therefore:
\begin{eqnarray}
B_1-B_2
&=& n \left [\ln 2\pi+ \ln s^2+n\right]+2 \ln n-n \left [\ln 2\pi+ \ln (s^2+m^2)+n\right] -\ln n\\
&=& n \left[ \ln s^2 - \ln (s^2+m^2) \right]+\ln n
\end{eqnarray}
Setting this equal to zero gives:
\begin{eqnarray}
n \left[ \ln s^2 - \ln (s^2+m^2) \right]+\ln n&=&0\\
\ln(s^2+m^2)-ln(s^2)&=&\frac{\ln n}{n}\\
\ln \left[s^2\left(\frac{m^2}{s^2}+1\right)\right]-ln(s^2)&=&\frac{\ln n}{n}\\
\ln \frac{m^2}{s^2}+1&=&\frac{\ln n}{n}\\
\frac{m^2}{s^2}+1&=&\mbox{exp} \left( \frac{\ln n}{n} \right)\\
r_c^2=\frac{m^2}{s^2}&=&\mbox{exp} \left( \frac{\ln n}{n} \right)-1\label{eqrc}
\end{eqnarray}

For $r$ below the critical value $r_c$ given by the above expression (equation~\ref{eqrc})
BIC favours the simpler model (ignore the ensemble mean, use the ensemble variance),
and for $r$ above this value it favours the model complex model (use the ensemble mean, use the ensemble variance).

As an example, for $n=10$ ({i.e. an ensemble of size 10}) this gives $r_c=\frac{m}{s}=0.509$.
Interestingly, this is at a lower point than the crossover between these models in the~\citet{jewsonh09c} study, which was 0.73.
This highlights the difference between model selection and monte carlo simulation methods. In this case using
model selection would lead to using the more complex model for smaller changes in predicted climate
than using the results of monte-carlo simulations would.

\begin{figure}[!ht]\begin{center}
\scalebox{0.8}{\includegraphics{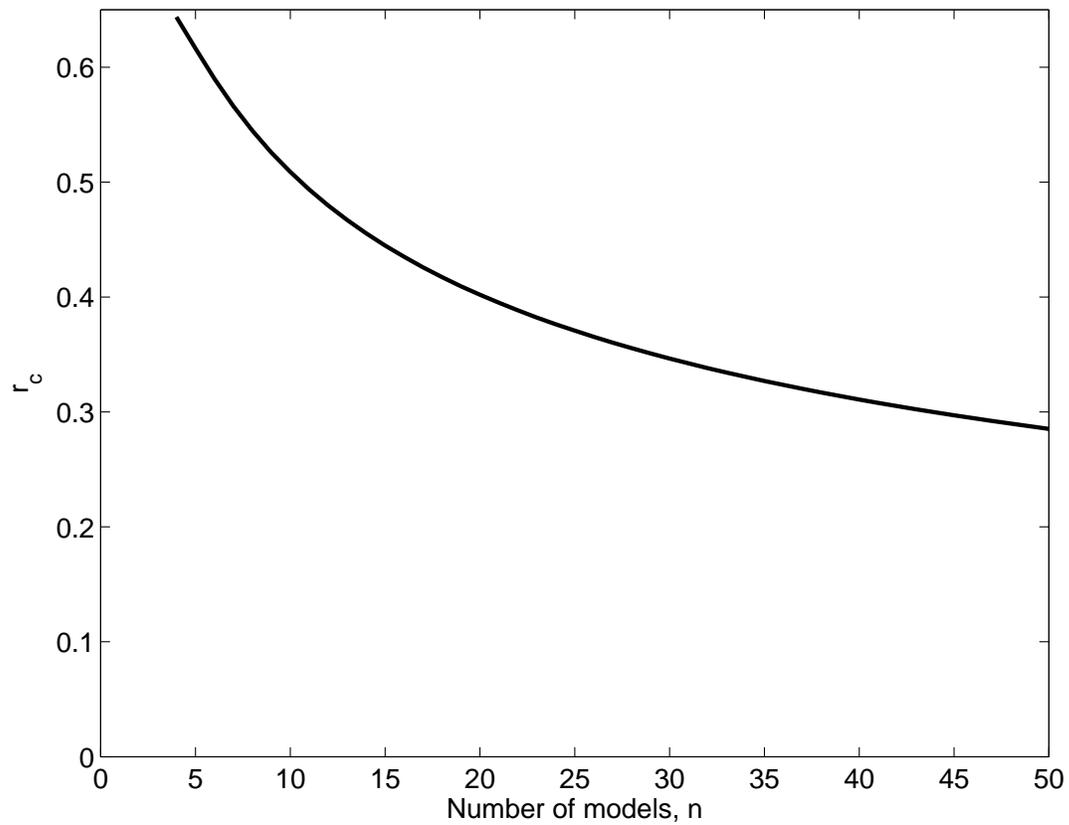}}
\end{center}\caption{
Critical values of $r$ for different ensemble sizes (where the models are independent, or where the variance across
the ensemble has been adjusted to compensate for any assumed dependency).
For values of $r$ greater than the critical value it is better (according to BIC) to use the ensemble mean than ignore it.
For values of $r$ less than the critical value it is better (according to BIC) to ignore the ensemble mean, and just use the
ensemble variance.
}
\label{fig1}\end{figure}

Figure~\ref{fig1} show the critical values of $r$ versus ensemble size, for a range of different ensemble sizes.

\section{Application to a Prediction of UK Rainfall}
\label{s5}

We now apply the ideas developed above to coupled climate model predictions for UK rainfall.
These predictions are based on multimodel ensembles generated during the CMIP project~\citep{cmip},
but with the ensemble variance inflated using the method described in~\citet{jewsonh09a},
using assumed correlations between the models of between $0$ and $0.75$.

\begin{figure}[!ht]\begin{center}
\scalebox{0.8}{\includegraphics{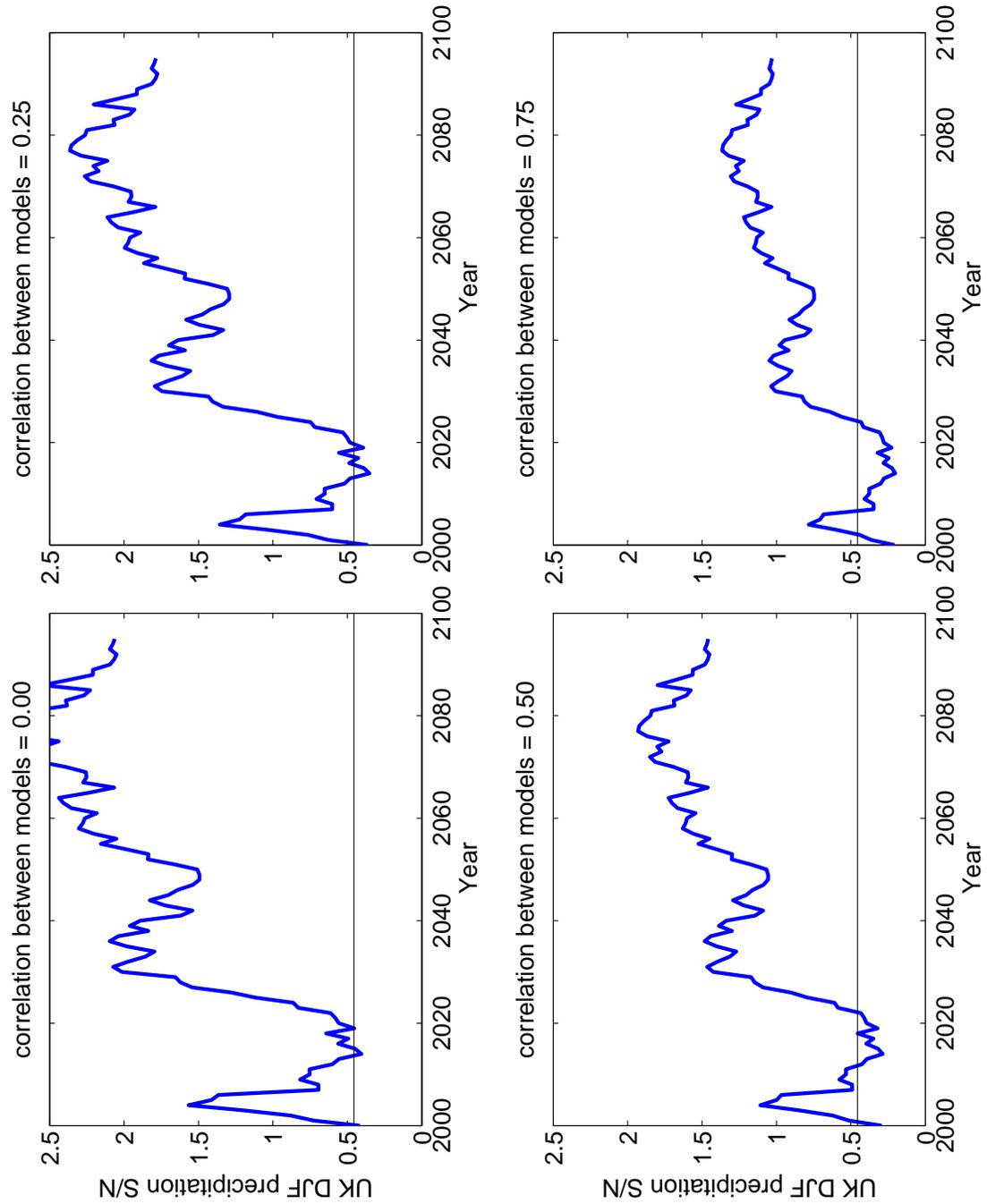}}
\end{center}\caption{
In blue, values of the prediction-to-uncertainty ratio $r$, for predictions of UK winter rainfall, for four different
assumptions about the correlations between the different models (0, 0.25, 0.5 and 0.75).
In black, the critical value of this ratio $r$, above which using the ensemble mean
gives better predictions than ignoring the ensemble mean.
}
\label{fig2}\end{figure}

In figure~\ref{fig2} we show the value of $r$ for these predictions versus lead time.
We also show the critical value of $r$ (which is $0.455$, since we have 14 models in the ensemble) as a horizontal line.
We see that, for a correlation of 0.75, the predictions only cross the critical value of $r$ in about 2025.
The implication is that before 2025 the spread across the models is so great than the ensemble mean is
poorly estimated to the extent that using it to derive the mean of a forecast gives a less good forecast than ignoring it
and use a mean change of zero.
It is clear, however, that the exact point at which the forecast crosses this threshold would be heavily influenced
by how the forecasts are smoothed in time.

\clearpage
\section{Summary}
\label{s6}

We have considered the question of how to make a prediction of future climate using output from a multimodel ensemble
of climate models. In particular we have considered what to do when the number of truly independent climate models
is small, and the spread between the model results is large, as is often the case for certain variables.
Using the standard BIC method for model selection, and making the perfect ensemble assumption, we have compared
using the ensemble mean and variance with just using the ensemble variance and setting the mean change in climate to zero.
This allows us to derive a simple expression for which of these two predictions to use.
We have then applied the expression to CMIP predictions for UK rainfall, and concluded that for those predictions
the ensemble mean should be ignored until at least 2025.

It is important to note that model selection is not the same as statistical testing.
In statistical testing one typically asks whether the data supports a certain hypothesis
(such as whether there are significant changes in UK rainfall),
and rejects the hypothesis unless the data is strongly consistent with the hypothesis.
In our case we are accepting the hypothesis that there is a change right from the start,
rather than testing that hypothesis.
We are then trying to decide, at a practical level, whether modelling the change is likely to give a better
forecast than ignoring the change.
If a predicted change is weak and uncertain, it can be perfectly logical to believe that there is a change
(based on independent arguments such as physical reasoning),
but that it is impossible to estimate it well enough for it to improve predictions.
That appears to be the case for changes in UK rainfall in the near future.
As an aside, it is also possible, for slightly stronger and/or less uncertain predictions, that it cannot be proven that a signal
is statistically significant, but that because you believe that the signal is real
(again, presumably based on independent reasoning)
then it is worthwhile to estimate the signal and include it in predictions.

There are a number of directions in which we plan to extend this work.
One is to try and include the `damped' predictions described in~\citet{jewsonh09c} in the model selection process. This
cannot be done using BIC, since the damped predictions are not maximum likelihood based. Still, there may be other methods
for model selection that one could use instead. The damped prediction methods will hopefully allow model predictions
with lower values of $r$ to become usable, and would give more accurate predictions for intermediate values of $r$.
Another direction would be to take an objective Bayesian approach to fitting the predictive distribution, in which case the normal
becomes a t distribution. Once again BIC can no longer be used, although presumably Bayes Factors, from which BIC is derived, can.

\bibliography{arxiv}

\begin{thebibliography}{5}
\providecommand{\natexlab}[1]{#1}
\providecommand{\url}[1]{\texttt{#1}}
\expandafter\ifx\csname urlstyle\endcsname\relax
  \providecommand{\doi}[1]{doi: #1}\else
  \providecommand{\doi}{doi: \begingroup \urlstyle{rm}\Url}\fi

\bibitem[Jewson and Hawkins(2009{\natexlab{a}})]{jewsonh09a}
S~Jewson and E~Hawkins.
\newblock {CMIP3 Ensemble Spread, Model Similarity, and Climate Prediction
  Uncertainty}.
\newblock \emph{arXiv:physics/0909.1890}, 2009{\natexlab{a}}.

\bibitem[Jewson and Hawkins(2009{\natexlab{b}})]{jewsonh09b}
S~Jewson and E~Hawkins.
\newblock {Improving the Expected Accuracy of Forecasts of Future Climate Using
  a Simple Bias-Variance Tradeoff}.
\newblock \emph{arXiv:physics/0911.1904}, 2009{\natexlab{b}}.

\bibitem[Jewson and Hawkins(2009{\natexlab{c}})]{jewsonh09c}
S~Jewson and E~Hawkins.
\newblock {Improving Uncertain Climate Forecasts Using a New Minimum Mean
  Square Error Estimator for the Mean of the Normal Distribution}.
\newblock \emph{arXiv:physics/0912.4395}, 2009{\natexlab{c}}.

\bibitem[Meehl et~al.(2007)Meehl, Covey, Delworth, Latif, McAvaney, Mitchell,
  Stouffer, and Taylor]{cmip}
G~Meehl, C~Covey, T~Delworth, M~Latif, B~McAvaney, J~Mitchell, R~Stouffer, and
  K~Taylor.
\newblock {The WCRP CMIP3 Multi-model Dataset: A New Era in Climate Change
  Research}.
\newblock \emph{Bulletin of the American Meteorological Society}, 88:\penalty0
  1383--1394, 2007.

\bibitem[Wasserman(2004)]{wasserman}
L~Wasserman.
\newblock \emph{{All of Statistics}}.
\newblock Springer, 2004.

\end{thebibliography}

\end{document}